\documentclass[twocolumn,aps,prl,epsfig,graphics]{revtex4}%
\usepackage{graphicx}
\usepackage{amsmath}
\usepackage{amsfonts}
\usepackage{amssymb}%
 
\begin{document}
\title{Atomic quantum dots coupled to BEC reservoirs}
\author{A. Recati,$^{1)}$ P.O. Fedichev,$^{1)}$ W. Zwerger$^{1)}$, J. von
Delft$^{3)}$, and P. Zoller$^{1)}$}

\address{$^{1}$ Institute for Theoretical Physics, University of Innsbruck,
A--6020 Innsbruck, Austria\\
$^{3}$ Sektion Physik, Universit\"{a}t M\"{u}nchen, Theresienstr. 37/III,
D-80333 M\"{u}nchen, Germany.}
\begin{abstract}
We study the dynamics of an atomic quantum dot, i.e. a single atom in a tight
optical trap which is coupled to a superfluid reservoir via laser transitions.
Quantum interference between the collisional interactions and the laser
induced coupling to the phase fluctuations of the condensate results in a
tunable coupling of the dot to a dissipative phonon bath, allowing an
essentially complete decoupling from the environment. Quantum dots embedded in
a 1D Luttinger liquid of cold bosonic atoms realize a spin-Boson model with
ohmic coupling, which exhibits a dissipative phase transition and allows to
directly measure atomic Luttinger parameters.

\end{abstract}
\maketitle

\address{$^{1}$ Institute for Theoretical Physics, University of Innsbruck,
A--6020 Innsbruck, Austria\\
$^{2}$ Dipartimento di Fisica, Universit\`{a} di Trento and INFM,
I-38050 Povo, Italy\\
$^{3}$ Sektion Physik, Universit\"{a}t M\"{u}nchen, Theresienstr. 37/III,
D-80333 M\"{u}nchen, Germany.}

A focused laser beam superimposed to a trap holding an atomic Bose-Einstein
condensate (BEC) \cite{BEC} allows the formation of an \emph{atomic quantum
dot} (AQD), i.e., a single atom in a tight trap \cite{tweezer,trap} which is
coupled to a reservoir of Bose-condensed atoms via laser transitions. This
configuration can be created, e.g., by spin-dependent optical potentials
\cite{spinlatt}, where atoms in the dot and the reservoir correspond to
different internal atomic states connected by Raman transitions. Atoms loaded
in the AQD will repel each other due to collisional interactions. In the limit
of strong repulsion, a collisional blockade regime can be realized where
either one or no atom occupies the dot, while higher occupations are excluded.
Below we will study the dynamics of such an AQD coupled to a BEC reservoir: as
the key feature we will identify the competition between two types of
interactions, namely the coupling of the atom in the dot to the BEC density
fluctuations via collisions, and the laser induced coupling to the fluctuating
condensate phase. Depending on the choice of interaction parameters, they can
interfere destructively or constructively, providing a tunable coupling of the
dot to the phonons in the condensate in the form of a spin-Boson model
\cite{spinboson}. In particular, an essentially complete decoupling of the dot
from the dissipative environment can be achieved, realizing a perfectly
coherent two-level system. This interference and tunability of the coupling of
the dot to the environment occurs for condensates in any dimensions. A
particularly interesting case is provided by a 1D superfluid
reservoir\cite{Petrov2000}, i.e., a bosonic Luttinger liquid of cold atoms
\cite{Haldane-Cazalilla}, where the system maps to a spin-Boson model with
ohmic coupling. The tunable dot-phonon coupling then allows the crossing of a
dissipative quantum phase transition \cite{spinboson,dpt}, and can serve also
as a novel spectroscopic tool to measure directly atomic Luttinger parameters.

\begin{figure}[ptb]
\includegraphics[scale=0.4]{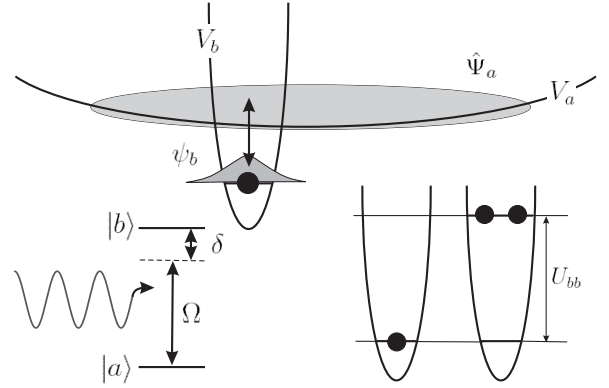}\caption{Schematic setup of an atomic quantum
dot coupled to a superfluid atomic reservoir. The Bose-liquid of atoms in
state $a$ is confined in a shallow trap $V_{a}(\mathbf{x})$. The atom in state
$b$ is localized in tightly confining potential $V_{b}(\mathbf{x})$. Atoms in
state $a$ and $b$ are coupled via a Raman transition with effective Rabi
frequency $\Omega$. A large onsite interaction $U_{bb}>0$ allows only a single
atom in the dot. }
\label{cap:setup}
\end{figure}

Let us consider cold bosonic atoms with two (hyperfine) ground
states $a$ and $b$ (Fig.~\ref{cap:setup}). Atoms in state $a$ form
a reservoir of atoms in a superfluid phase, held in a shallow
trapping potential $V_{a}(\mathbf{x})$. The AQD is formed by
trapping atoms in state $b$ in a tightly confining potential
$V_{b}(\mathbf{x})$ produced, e.g., by a focused laser beam
induced potential or by a deep optical lattice potential which is
only seen by atoms in state $b$ \cite{spinlatt}. Within the
standard pseudopotential description, the collisional interaction
of atoms in the two internal levels $\alpha ,\beta=a,b$ is
described by a set of coupling parameters $g_{\alpha\beta}=4\pi
a_{\alpha\beta}\hbar^{2}/m$ with scattering lengths
$a_{\alpha\beta}$ \cite{Tightconfinement} and atomic mass $m$. We
assume that the reservoir atoms are coupled via a Raman transition
to the lowest vibrational state in the AQD, where spontaneous
emission is suppressed by a large detuning from the excited
electronic states. Thus, following arguments analogous to those in
the derivation of the Bose-Hubbard model of cold atoms in an
optical lattice \cite{spinlatt}, we obtain an effective
Hamiltonian,
\begin{align}\label{eq:bHam}
H_{b}+H_{ab} &  =\left(  -\hbar\delta_{0}+g_{ab}\int
d\mathbf{x~|}\psi
_{b}(\mathbf{x})|^{2}\hat{\rho}_{a}(\mathbf{x})\right)
\hat{b}^{\dagger}
\hat{b}\\
&
+\frac{U_{bb}}{2}\hat{b}^{\dagger}\hat{b}^{\dagger}\hat{b}\hat{b}+\int
d\mathbf{x}\hbar\Omega(\hat{\Psi}_{a}(\mathbf{x})\psi_{b}(\mathbf{x})\hat
{b}^{\dagger}+\mathrm{h.c.})\nonumber,
\end{align}
Here $\hat{\Psi}_{a}(\mathbf{x})$ is the annihilation operator for an atom $a$
at the point $\mathbf{x}$, while $\hat{\rho}_{a}(\mathbf{x})=\hat{\Psi}%
_{a}^{\dagger}(\mathbf{x})\hat{\Psi}_{a}(\mathbf{x})$ is the associated
operator for the density. The operator $\hat{b}$ destroys a $b$-atom in the
dot in the lowest vibrational state with wave function $\psi_{b}(\mathbf{x})$.
The first term in Eq.~(\ref{eq:bHam}) includes the Raman detuning $\delta_{0}%
$, and the collisional interactions between the $b$-atoms with the
reservoir. The second term describes the onsite repulsion
$U_{bb}\sim g_{bb}/l_{b}^{3}>0$ between atoms in the dot with
$l_{b}$ the size of the ground state wave function. The last term
in (\ref{eq:bHam}) is the laser induced coupling between $a$ and
$b$ atoms with effective Rabifrequency $\Omega$. In writing
(\ref{eq:bHam}) we exclude coupling to higher vibrational states
in the dot, assuming that these states are  offresonant
\cite{spinlatt}.

At sufficiently low temperatures the reservoir atoms in $a$ form a superfluid
Bose liquid with an equilibrium liquid density $\rho_{a}$. The only available
excitations at low energies are then phonons with linear dispersion
$\omega=v_{s}|\mathbf{q}|$ and sound velocity $v_{s}$. With the assumption
that the number of condensate atoms inside the dot is much larger than one,
$n_{a}=\rho_{a}l_{b}^{3}>>1$, i.e. $l_{b}$ is much larger than the average
interparticle spacing in the BEC reservoir, the quantum dot is coupled to a
coherent matter wave and the Bose-field operator can be split into magnitude
and phase, $\hat{\Psi}_{a}(\mathbf{x})\sim\hat{\rho}_{a}(\mathbf{x}%
)^{1/2}e^{-i\hat{\phi}(\mathbf{x})}$. The dimensionless proportionality
coefficient is the bare condensate fraction, which depends on non-universal
short distance properties of the Bose-liquid. This representation does not
require the existence of a true condensate and can be also used to describe
both 2D and 1D superfluid systems \cite{Popov}. The density operator can be
expressed in terms of the density fluctuation operator $\hat{\Pi}$: $\hat
{\rho}_{a}(\mathbf{x})=\rho_{a}+\hat{\Pi}(\mathbf{x})$, which is canonically
conjugate to the superfluid phase $\hat{\phi}$. In the long wavelength
approximation the dynamics of the superfluid is described by a (quantum)
hydrodynamic Hamiltonian \cite{Popov}
\begin{equation}
H_{a}=\frac{1}{2}\int d\mathbf{x}\left(  \frac{\hbar^{2}}{m} \rho_{s}%
|\nabla\hat{\phi}|^{2}(\mathbf{x})+\frac{mv_{s}^{2}}{\rho_{a}} \hat{\Pi}%
^{2}(\mathbf{x})\right)  , \label{eq:Hah}%
\end{equation}
where $\rho_{s}$ is the density of the superfluid fraction (at zero
temperature $\rho_{s}=\rho_{a}$). The quadratic Hamiltonian (\ref{eq:Hah}) is
easily diagonalized by introducing standard phonon operators $b_{\mathbf{q}}$
via the following transformation:
\begin{equation}
\hat{\phi}(\mathbf{x})=i\sum_{\mathbf{q}} \left|  \frac{m v_{s}}%
{2\hbar\mathbf{q}V\rho_{a}}\right|  ^{1/2} e^{i\mathbf{q\cdot x}%
}(b_{\mathbf{q}}-b_{-\mathbf{q}}^{\dagger}) \label{eq:phidef}%
\end{equation}
\begin{equation}
\hat{\Pi}(\mathbf{x})=\sum_{\mathbf{q}}\left|  \frac{\hbar\rho_{a}\mathbf{q}}
{2v_{s}Vm}\right|  ^{1/2}e^{i\mathbf{q\cdot x}}(b_{\mathbf{q}}+b_{-\mathbf{q}
}^{\dagger}) \label{eq:pi}%
\end{equation}
with $V$ the sample volume. Accordingly, the Hamiltonian (\ref{eq:Hah}) takes
the form of a collection of harmonic sound modes: $H_{a}=\hbar v_{s}%
\sum_{\mathbf{q}}|\mathbf{q}|b_{\mathbf{q}}^{\dagger}b_{\mathbf{q}}$. Since
the excitations of a weakly interacting Bose-liquid are phonon-like only for
wavelengths larger than the healing length $\xi\geq l_{b}$, the summation over
the phonon modes is cutoff at a frequency $\omega_{c}=v_{s}/\xi\approx
g_{aa}\rho_{a}/\hbar$.

In the following we consider the collisional blockade limit of large onsite
interaction $U_{bb}$, where only states with occupation $n_{b}=0$ and $1$ in
the dot participate in the dynamics, while higher occupations are suppressed
by the large collisional shift. This situation and its description is
analogous to the Mott insulator limit in optical lattices
\cite{spinlatt,Mott,dipoleblock}. The requirements are that the Raman detuning
and Rabi couplings are much smaller than $U_{bb}/\hbar$, which provides a
small parameter to eliminate the states with higher occupation numbers
perturbatively. As discussed below, a Feshbach resonance can help in achieving
this large $U_{bb}$ limit \cite{Feshbach}. Thus, the quantum state of a dot is
described by a pseudo-spin-$1/2$, with the spin-up or spin-down state
corresponding to occupation by a single or no atom in the dot. Using standard
Pauli matrix notation, the dot occupation operator $\hat{b}^{\dagger}\hat{b}$
is then replaced by $(1+\sigma_{z})/2$ while $\hat{b}^{\dagger}\rightarrow
\sigma_{+}$. Furthermore, the dominant coupling between the AQD and the
superfluid reservoir arises from the long wavelength phonons. For wavevectors
$|\mathbf{q}|l_{b}\ll1$, the phonon field operators may be replaced by their
values at $\mathbf{x}=\mathbf{0}$. Neglecting the density fluctuations in the
Raman coupling (see below) and an irrelevant constant, the Hamiltonian
(\ref{eq:bHam}) is simplified to
\begin{equation}
\label{eq:Hbnew}H_{b}+H_{ab}=\left(  -\frac{\hbar\delta}{2}+\frac{g_{ab}}%
{2}\hat{\Pi}(\mathbf{0})\right)  \sigma_{z}+\frac{\hbar\Delta}{2}(\sigma
_{+}e^{-i\hat{\phi}(\mathbf{0})}+\mathrm{h.c.}).
\end{equation}
Here $\Delta\sim\Omega n_{a}^{1/2}$ is an effective Rabi frequency with
$n_{a}^{1/2}$ the bosonic enhancement factor due to the BEC reservoir, and
with a proportionality coefficient which depends on the bare condensate
fraction and the explicit form of the wavefunction $\psi_{b}$. The form of the
Rabi coupling in (\ref{eq:Hbnew}) only applies for $\Delta\ll \omega_c$, 
which is the interesting regime in the spin-Boson model discussed below.
In (\ref{eq:Hbnew}) the detuning has been renormalized to include a mean field shift and a shift
due to the virtual admixture of the double occupied state in the dot,
$-\hbar\delta_{0}+g_{ab}\rho_{a}+(\hbar\Delta)^{2}/(-2U_{bb}%
)\equiv-\hbar\delta$.

Moreover, the validity of the above model requires a strong collisional
interaction of atoms in the AQD, $g_{bb}\gg g_{aa}$. This follows
from the inequalities $n_{a}\gg1$, $\Delta\ll \omega_{c}$ and 
the single
occupancy condition
$\hbar\Delta\ll U_{bb}$. A magnetic or (Raman laser induced) optical
Feshbach resonance in the $b$ channel will assist in achieving
this limit \cite{Feshbach}. These resonances arise from coupling
to a bound\ molecular state in an energetically closed collisional
channel. In the case of an optical Feshbach resonance, for
example, the onsite interaction due to the laser induced Raman
coupling of two $b$ atoms in the dot to a molecular bound state
leads in Eq.~(\ref{eq:bHam}) to the replacement $U_{bb}\rightarrow
U_{bb}^{\mathrm{(res)}}=U_{bb}+g^{2}/\delta_{m}$, where the second
term describes the resonant enhancement with $g$ an effective
Raman Rabi frequency and $\delta_{m}$ the detuning from the
molecular resonance is (valid for $g<|\delta_{m}|$). A finite
lifetime of the molecular state, e.g.
due to collisions with $a$ atoms, introduces a width $\delta_{m}%
\rightarrow\delta_{m}-i\Gamma/2$ \cite{Petrov}. Thus for detunings
$|\delta_{m}|\gg\Gamma$ we have $U_{bb}\rightarrow U_{bb}^{\mathrm{(res)}%
}+i\gamma_m/2$ with $\gamma_m=(g^{2}/\delta_{m}^{2})\Gamma\ll U_{bb}%
^{\mathrm{(res)}}$. Returning to the Hamiltonian (\ref{eq:Hbnew})
we see that besides the resonantly enhanced onsite interaction we
have a non-Hermitian loss term
$\sim(\hbar\Delta/U_{bb}^{\mathrm{(res)}})^{2}\gamma_m$ which is
strongly suppressed in the collisional blockade limit.

Eventually after a unitary transformation $H=S^{-1}(H_{a}+H_{b}+H_{ab})S$ with
$S=\exp(-\sigma_{z}i\hat{\phi}(\mathbf{0})/2)$ the dynamics of the AQD coupled
to the phonons of the superfluid reservoir is described by a spin-Boson type
Hamiltonian \cite{spinboson}
\begin{equation}
H=-\frac{\hbar\Delta}{2}\sigma_{x}+\sum_{\mathbf{q}}\hbar\omega_{\mathbf{q}%
}b_{\mathbf{q}}^{\dagger}b_{\mathbf{q}}+\left(  -\delta+\sum_{\mathbf{q}%
}\lambda_{\mathbf{q}}(b_{\mathbf{q}}+b_{\mathbf{q}}^{\dagger})\right)
\frac{\hbar\sigma_{z}}{2}\label{eq:Hspinboson}%
\end{equation}
The coherence in the reservoir is crucial for the most important feature of
our cold atom version of the spin-Boson model (\ref{eq:Hspinboson}), namely
the fact that the collisional interactions and those arising from the coupling
of the Rabi term to the condensate phase add coherently in the amplitudes 
of the total phonon coupling 
\begin{equation}
\lambda_{\mathbf{q}}=\left\vert \frac{m\hbar\mathbf{q}v_{s}^{3}}{2V\rho_{a}%
}\right\vert ^{1/2}\left(  \frac{g_{ab}\rho_{a}}{mv_{s}^{2}}-1\right)
.\label{eq:lamgas}%
\end{equation}
In particular, we see that for a repulsive inter-species
interaction, $g_{ab}>0$, both contributions interfere destructively: the
effect of the phonon excitation in a laser-driven transition $a\leftrightarrow
b$ can be precisely cancelled by the change in the direct \textquotedblleft
elastic\textquotedblright\ interaction between the liquid and the $b$ atoms,
as described by the first term in Eq.~(\ref{eq:Hbnew}). In a weakly
interacting gas $mv_{s}^{2}=\rho_{a}g_{aa}$ and thus the coupling constants
$\lambda_{\mathbf{q}}$ vanish at $g_{ab}=g_{aa}$. At this special point, we
have thus formed the analogue of a \textquotedblleft charge\textquotedblright%
\ qubit of solid state physics \cite{JJ}, i.e., a coherent superposition of
occupation and non-occupation of the dot, which - at vanishing detuning -
exhibits perfect Rabi oscillations of the AQD's occupancy. Notice that
the simple form (\ref{eq:lamgas}) is valid only for $|\mathbf{q}|l_{b}\ll 1$, 
while for larger values of the momentum the interaction starts to decrease 
proportional to the Fourier transform of the wavefunction $\psi_{b}$.

We discuss now the importance of the neglected density fluctuations in the
Rabi term and their effect at the decoupling point. After the unitary
transformation above, these fluctuations give rise to a perturbation
$H^{\prime}=\frac{\hbar\Delta}{2}\frac{\hat{\Pi}(\mathbf{0})}{2\rho_{a}}
\sigma_{x}\label{Vprime}$%
which is small compared to the collisional interaction in
Eq.(\ref{eq:Hbnew}) provided that $\hbar\Delta\ll\rho_{a}g_{ab}$.
Now unless $g_{ab}$ is much smaller than $g_{aa}$, this condition
is precisely equivalent to the condition $\Delta\ll\omega_{c}$
discussed above. In particular, if we are at the decoupling point,
the only remaining interaction with the phonon bath is given by
$H^{\prime}$. For $\delta=0$ the Hamiltonian can then be rewritten
as
\begin{equation}
H=-\frac{\hbar\Delta}{2}\left(  1-\frac{\Pi(\mathbf{0})}{2\rho_{a}}\right)
\sigma_{x}.\label{HiB}%
\end{equation}
This is an independent Boson model which can be diagonalized exactly
\cite{Mahan}. Since the phonons now no longer couple to the $b$ atom
occupation $\sigma_{z}$ and, moreover, the coupling constants are proportional
to the small parameter $\Delta/\omega_{c}$, one obtains perfect Rabi
oscillations except for a small reduction in amplitude by a factor
$\exp{-\gamma(\Delta/\omega_{c})^{2}}\approx1$, where $\gamma\alt(\rho
_{a}a_{aa}^{3})^{1/2}\ll1$ is proportional to the small gas parameter.

Let us now turn to discuss the properties of system when it can be described
by the Hamiltonian Eq. (\ref{eq:Hspinboson}). The system is characterized by
the effective density of states
\begin{equation}
J(\omega)=\sum_{\mathbf{q}}\lambda_{\mathbf{q}}^{2}\delta(\omega
-\omega_{\mathbf{q}})=2\alpha\omega^{s}, \label{eq:Jdef}%
\end{equation}
where $\alpha\sim(g_{ab}\rho_{a}/mv_{s}^{s}-1)^{2}$ is the dissipation
strength due to the spin-phonon coupling and $D=s$ the dimension of the
superfluid reservoir. In the standard terminology \cite{spinboson}, $s=1$ and
$s>1$ correspond to the ohmic and superohmic cases, respectively. In the
superohmic case, the resulting dynamics of the AQD is a damped oscillation at
vanishing detuning $\delta=0$, consistent with the result of simple Bloch
equation analysis. It may be observed by following the population of the atoms
$b$ in the presence of the laser coupling. The associated frequency
$\tilde{\Delta}$ and damping $\Gamma$ can also be obtained by measuring a weak
field absorption spectrum, which would exhibit the oscillations with frequency
$\tilde{\Delta}$ as a line splitting, and $\Gamma$ as the linewidth.

\begin{figure}[ptb]
\includegraphics[scale=0.42]{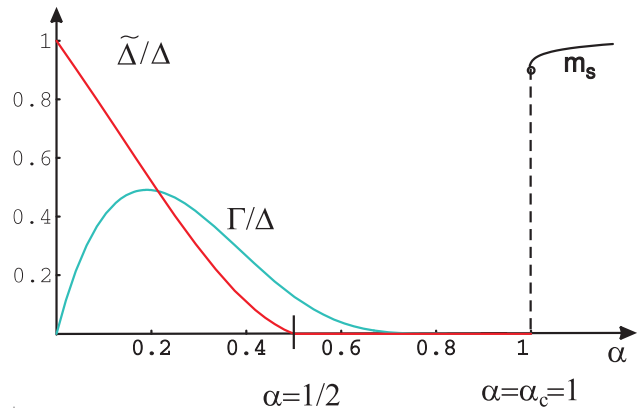}\caption{The oscillation frequency
$\tilde{\Delta}$ and the damping rate $\Gamma$ as functions of the coupling
strength $\alpha$. For the damping in the range $1/2<\alpha <1$ we used the
approximate expression $\Gamma=(1/2- \alpha)\tan (\pi\alpha) \Delta_r(\alpha)$ \cite{WZ83}. The dissipative phase transition shows up as a jump of size
$m_{s}$ in the occupation as the detuning $\delta$ is changed from small
negative to positive values. }
\label{fig:1DSB}
\end{figure}

A much richer dynamics appears for ohmic dissipation (Fig. \ref{fig:1DSB}).
In this case, the system
exhibits a zero temperature dissipative phase transition, as a
function of the dissipation strength, at a critical value 
$\alpha_{c}=1$ for $\Delta\ll \omega_c$  \cite{spinboson}. 
In the symmetry broken regime $\alpha>\alpha_c$ the occupation probability of 
the $b$ atom will exhibit  a finite jump from $(1-m_s)/2$ to $(1+m_s)/2$
as the detuning $\delta$ is changed across zero. The spontaneous polarization
$m_s$ is a function of $\alpha$ approaching $m_s\simeq 0.9$ for $\alpha\rightarrow
\alpha_c^+$ \cite{Anderson} and $m_s=1$ for $\alpha\gg\alpha_c$. As a result the $b$
occupation probability is almost unity for any $\alpha>\alpha_c$.
Instead in the regime $\alpha<\alpha_c$, at vanishing detuning, 
the average population of the $b$ atoms is $1/2$.
In particular, for $\alpha<1/2$ one has
damped Rabi oscillations. In terms of the characteristic
frequency scale $\Delta_r(\alpha)=\Delta(\Delta/\omega _{c})^{\alpha/(1-\alpha)}$,
the effective Rabi oscillation frequency $\tilde{\Delta}$ and
damping rate $\Gamma$ are given by 
$\tilde{\Delta}=\cos{\eta}\cdot\Delta_r(\alpha)$ and 
$\Gamma=\sin{\eta}\cdot\Delta_r(\alpha)$ with $\eta=\pi\alpha/(2(1-\alpha))$ \cite{spinboson}.
This result holds as long as $T\alt
T_*=\hbar\Delta_{r}/\alpha$.  At higher temperatures the
dynamics is incoherent and no oscillations should be visible.
For $1/2<\alpha<1$ the Rabi oscillations disappear 
and the behaviour is completely incoherent \cite{diss-qft}. Only numerical
results are available for the dynamics in this regime
\cite{Hofstetter}.

The ohmic spin-Boson model is achieved by embedding the AQD in an atomic
quantum wire as realized recently in \cite{EsslingerBloch}. For transverse
harmonic trapping both the temperature $T$ and the chemical potential need to
be smaller than the frequency of the transverse confinement $\omega_{\perp}$.
The proper description of a 1D superfluid in terms of the hydrodynamics
Hamiltonian (\ref{eq:Hah}) is based on the Haldane-Luttinger approach
\cite{Haldane-Cazalilla}. The definitions of the phase and the density
fluctuations in terms of the phonon operators are given by
Eqs.(\ref{eq:phidef},\ref{eq:pi}) with $v_{s}=\hbar\pi\bar{\rho}_{a}/(mK)$,
where $\bar{\rho}_{a}\sim\rho_{a}l_{\perp}^{2}$ is the 1D density of the cloud
and $l_{\perp}=(\hbar/m\omega_{\perp})^{-1/2}$ the transverse ground state
size. The Luttinger parameter $K$ depends on the density of the liquid through
the combination $\gamma_{a}=m\bar{g}_{aa}/\hbar^{2}\bar{\rho}_{a}$. The
corresponding 1D interaction constant $\bar{g}_{aa}$ is related to the 3D
scattering length $a_{aa}$ \cite{Olshanii} and $\bar{g}_{aa}=2\hbar
\omega_{\perp}a_{aa}$ as long as $a_{aa}\ll l_{\perp}$.

Using Eq.(\ref{eq:lamgas}) for the coupling constants $\lambda_{q}$, we find
\begin{equation}
\alpha=\frac{1}{8K(\gamma_{aa})}\left(  \frac{\gamma_{ab}K(\gamma_{aa})^{2}%
}{\pi^{2}}-1\right)  ^{2}%
\end{equation}
where $\gamma_{ab}=m\bar{g}_{ab}/\hbar^{2}\bar{\rho}_{a}$. Similar to the
discussion above, the interference between the direct interaction and the
Rabi-transition terms in the Hamiltonian (\ref{eq:Hbnew}) leads to the
disappearance of AQD-phonon coupling at $\gamma_{ab}=\pi^{2}/K^{2}$. The
Luttinger liquid parameter $K$ can thus be determined by tuning the AQD to the
decoupling point via a change in the known interaction constant $\gamma_{ab}$.
Thus, observation of the dynamics of the dot coupled to a Luttinger liquid
provides a novel tool to measure $K$ directly. For a weakly interacting 1D
liquid $\gamma_{aa}\ll1$ the Bogoliubov approximation applies, giving
$K=\pi/\sqrt{\gamma_{aa}}$. In recent experiments \cite{EsslingerBloch} values
$\gamma_{aa}\geq1$ have been reached in in an array of independent
one-dimensional tubes . In this case the critical value $\alpha_{c}=1$ for the
dissipative phase transition is reached at $a_{ab}/a_{aa}\sim1$.

In conclusion, we have shown that an AQD coupled to a superfluid
reservoir leads to spin-Boson model with tunable parameters.  The
present model may be readily extended to arrays of AQDs, where the
pseudo-spins can interact with \textquotedblleft
host\textquotedblright\ liquid in a collective manner.

Discussions with M. Cazalilla, J.I. Cirac, and S. Kehrein are
acknowledged. Work supported in part by the Austrian Science
Foundation, the Institute for Quantum Information and the European
Commission RTN Network Contract No. HPRN-CT-2000-00125.

\end{document}